\documentclass[twocolumn]{article}
\usepackage{flushend}
\usepackage{waspaa23,amsmath,graphicx,url,times}
\usepackage{caption}
\usepackage{algorithm}
\usepackage{algpseudocode}
\usepackage{subcaption}
\usepackage{amssymb}
\usepackage{tabularx}
\usepackage{longtable}
\usepackage{multirow}
\usepackage[tableposition=top]{caption}
\usepackage[table, svgnames, dvipsnames]{xcolor}
\usepackage{color,soul}
\usepackage{hyperref}
\usepackage{makecell, cellspace, caption}
\makeatletter
\def\blfootnote{\xdef\@thefnmark{}\@footnotetext}
\makeatother

\title{Histogram Layer Time Delay Neural Networks for Passive Sonar Classification}

\name{Jarin Ritu$^{1}$,
      Ethan Barnes$^{1}$,
      Riley Martell$^{2}$,
      Alexandra Van Dine$^{2}$,
      Joshua Peeples$^{1}$\thanks{DISTRIBUTION STATEMENT A. Approved for public release. Distribution is unlimited. This material is based upon work supported by the Under Secretary of Defense for Research and Engineering under Air Force Contract No. FA8702-15-D-0001. Any opinions, findings, conclusions or recommendations expressed in this material are those of the author(s) and do not necessarily reflect the views of the Under Secretary of Defense for Research and Engineering. \textsuperscript{\textcopyright} 2023 Massachusetts Institute of Technology. Delivered to the U.S. Government with Unlimited Rights, as defined in DFARS Part 252.227-7013 or 7014 (Feb 2014). Notwithstanding any copyright notice, U.S. Government rights in this work are defined by DFARS 252.227-7013 or DFARS 252.227-7014 as detailed above. Use of this work other than as specifically authorized by the U.S. Government may violate any copyrights that exist in this work.
      \url{https://github.com/Peeples-Lab/HLTDNN}}}
\address{$^1$Department of Electrical and Computer  Engineering, Texas A\&M University, College Station, TX, USA \\ $^2$Massachusetts Institute of Technology Lincoln Laboratory, Lexington, MA, USA\\
}

\begin{document}
\ninept
\maketitle
 
\begin{abstract}
Underwater acoustic target detection in remote marine sensing operations is challenging due to complex sound wave propagation. Despite the availability of reliable sonar systems, target recognition remains a difficult problem. Various methods address improved target recognition. However, most struggle to disentangle the high-dimensional, non-linear patterns in the observed target recordings. In this work, a novel method combines a time delay neural network and histogram layer to incorporate statistical contexts for improved feature learning and underwater acoustic target classification.  The proposed method outperforms the baseline model, demonstrating the utility in incorporating statistical contexts for passive sonar target recognition. The code for this work is publicly available. 

\end{abstract}


\section{Introduction} \label{sect:introduction}
\begin{figure}[t]
        \centering
        \includegraphics[width=.48\textwidth]{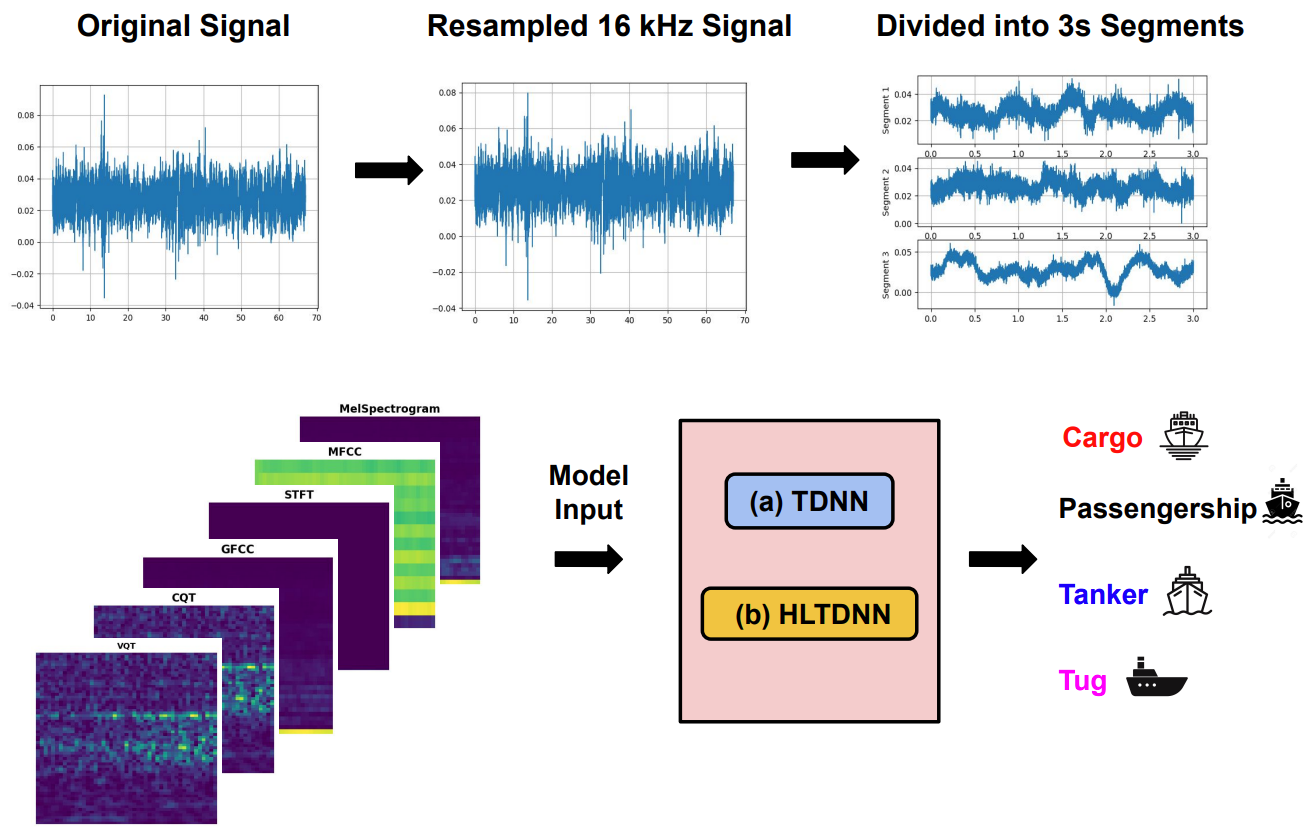}
        \caption{Overall experimental work flow. Each signal is resampled to $16$ kHz and binned into three second segments. After dividing the signals and corresponding segments into training, validation, and test partitions, several time-frequency features are extracted. The features are then passed into the model and classified as one of the four vessel types.} 
        \label{fig:Fig1_WorkFlow}
    \end{figure}
Underwater acoustic target recognition (UATR) technology plays a crucial role in a variety of domains, including biology \cite{beckler2022multilabel}, carrying out search and rescue operations, enhancing port security \cite{kita2022passive}, and mapping the ocean floor \cite{mellema2006reverse}. One of the primary target detection techniques used by modern crafts, such as unmanned underwater vehicles, is passive sonar \cite{cheng2022uuv}. Passive sonar is an underwater acoustic technology that uses hydrophones to detect and analyze sound waves in the ocean \cite{de2022passive}. Unlike active sonar, passive sonar resolves targets from the natural sounds of the ocean and the noises produced by ships and other underwater vehicles. Processing and analyzing passive sonar data can be challenging due to the high volume of data and environmental complexity \cite{neupane2020review}. Signal processing techniques are often used to analyze ship-generated noise such as low frequency analysis and recording (LOFAR) spectra \cite{luo2023survey}. The Detection of Envelope Modulation on Noise (DEMON) is an approach that has been successfully used for target detection and recognition in passive sonar \cite{4318809,hashmi2023novel,ambat2022performance}. Despite their success, these approaches use handcrafted features that can be difficult to extract without domain expertise \cite{cao2016deep}.

\begin{figure*}[htb]
    \centering
    \includegraphics[width=.92\linewidth]{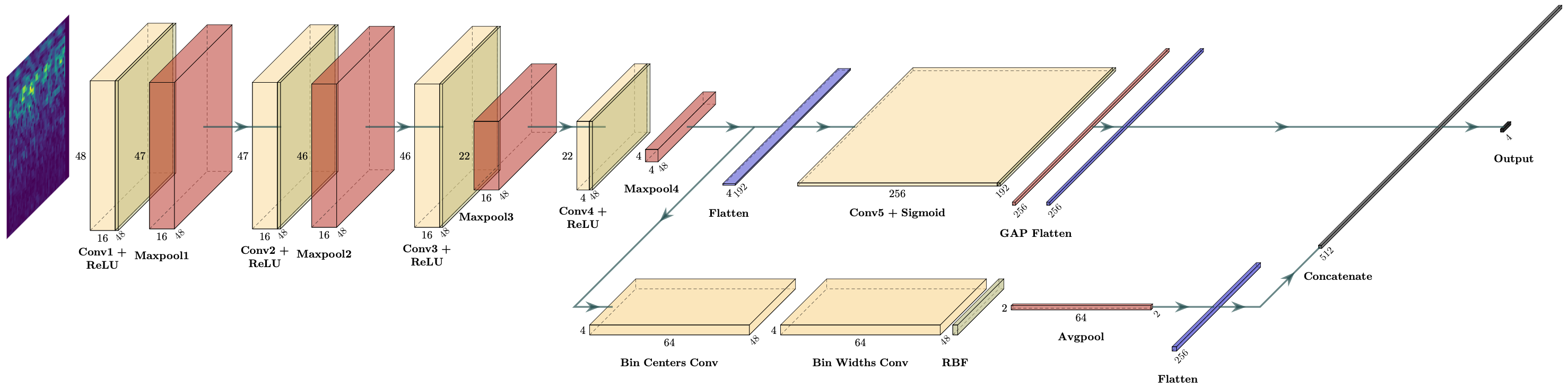}
    \caption{Proposed HLTDNN architecture. The histogram layer is added in the parallel with the baseline TDNN model through the bin center and width convolution layers with the radial basis activation function (RBF) and average pooling layer.}
    \label{fig:enter-label}
\end{figure*}

 Artificial neural networks (ANNs), such as convolutional neural networks (CNNs) and time delay neural networks (TDNNs), provide an end-to-end process for automated feature learning and follow-on tasks (\textit{e.g.}, detection and classification of signals) \cite{jing1994multilayered,ashok2022improving,irfan2021deepship,doan2020underwater}. The TDNN has shown success in simulating long-term temporal dependencies \cite{peddinti2015time} and can be modeled as a 1D CNN \cite{ashok2022improving}. Thus, the TDNN can adaptively learn the sequential hierarchies of features, but does not explicitly account for the statistics of passive sonar data. These are difficult to model for feature extraction \cite{komari2018passive,trevorrow2021examination}. The statistics of the signals can describe the acoustic texture of the targets of interest \cite{trevorrow2021examination}. Texture generally falls into two categories: statistical and structural \cite{srinivasan2008statistical,ji2022structural,zhu2021learning,peeples2021histogram}.Statistical context in audio analysis involves studying the amplitude information of the audio signal. One way to capture amplitude information is by using probability density functions {\cite{trevorrow2021examination}}.
However, traditional artificial neural network (ANN) approaches, like convolutional neural networks (CNNs) and time-delay neural networks (TDNNs), have shown a bias towards capturing structural textures rather than statistical texture \cite{ji2022structural,zhu2021learning,peeples2021histogram}. This bias limits their ability to directly model the statistical information required to capture acoustic textures accurately. To overcome this shortcoming, histogram layers can be integrated into ANNs to incorporate statistical context \cite{peeples2021histogram}. Methods that combine both structural and statistical textures have improved performance for other tasks such as image classification and segmentation \cite{ji2022structural,zhu2021learning,peeples2021histogram}. In this work, we propose a new TDNN architecture that integrates histogram layers for improved target classification. Our proposed workflow is summarized in Figure \ref{fig:Fig1_WorkFlow}. The contributions of this work are as follows:

\begin{itemize}
    \item Novel TDNN architecture with histogram layer (HLTDNN) for passive sonar target classification
    \item In-depth qualitative and quantitative comparisons of TDNN and HLTDNN across a suite of time-frequency features.
\end{itemize}

\section{Method}
\subsection{Baseline TDNN Architecture}
The TDNN architecture consisted of several convolution layers with the ReLU activation function and max pooling. 2D convolutional features were extracted from the time-frequency input to capture local relationships between the vessel's frequency information \cite{thienpondt2021integrating}. {Padding was added to the input time-frequency feature to maintain the spatial dimensions of the resulting features maps.} After each convolution operation and ReLU activation function, the features were pooled along the time axis with desired kernel length $L$ (\textit{e.g.}, max pooling kernel of size $1 \times L$) to aggregate the feature information while maintaining the temporal dependencies similar to other TDNNs \cite{peddinti2015time,thienpondt2021integrating}. After the fourth convolutional block, the features are flattened and then passed through a final 1D convolutional layer followed by a sigmoid activation function and global average pooling layer (GAP). 


\subsection{Proposed HLTDNN}
The baseline TDNN is focused on the ``structural" (\textit{e.g.}, local) acoustic textures of time and frequency as well as the temporal dependencies in the data. However, the model does not directly consider the statistical aspects of the data. A histogram layer \cite{peeples2021histogram} can be added in parallel to the baseline TDNN model to capture statistical features to assist in improving classification performance.  Given input features, $\mathbf{X} \in \mathbb{R}^{M \times N \times D}$, where $M$ and $N$ are the spatial (or time-frequency) dimensions while $D$ is the feature dimensionality, the output tensor of the local histogram layer with $B$ bins, $\mathbf{Y} \in \mathbb{R}^{R \times C \times B \times D}$ with spatial dimensions $R$ and $C$ after applying a histogram layer with kernel size $S \times T$ is shown in (\ref{eqn:binning}):
\begin{equation}
    Y_{rcbd} =  
	\cfrac{1}{ST}\sum_{s=1}^{S}\sum_{t=1}^{T}e^{-\gamma_{bd}^2\left(x_{r+s,c+t,d}-\mu_{bd}\right)^2}
	\label{eqn:binning}
\end{equation}
where the bin centers ($\mu_{bd}$) and bin widths ($\gamma_{bd}$)  of the histogram layer are learnable parameters. Each input feature dimension is treated independently, resulting in $BD$ output histogram feature maps. The histogram layer takes input features and outputs the ``vote" for a value in the range of $[0, 1]$. The histogram layer can be modeled using convolution and average pooling layers as shown in Figure \ref{fig:enter-label}. Following previous work \cite{peeples2021histogram}, the histogram layer is added after the fourth convolutional block (\textit{i.e.}, convolution, ReLU, and max pooling) and its features are concatenated with the TDNN features before the final output layer. 

\section{Experimental Procedure}
\label{sec:exp}
\subsection{Dataset Description}
\begin{figure}[H]
    \centering
    \includegraphics[scale=.33]{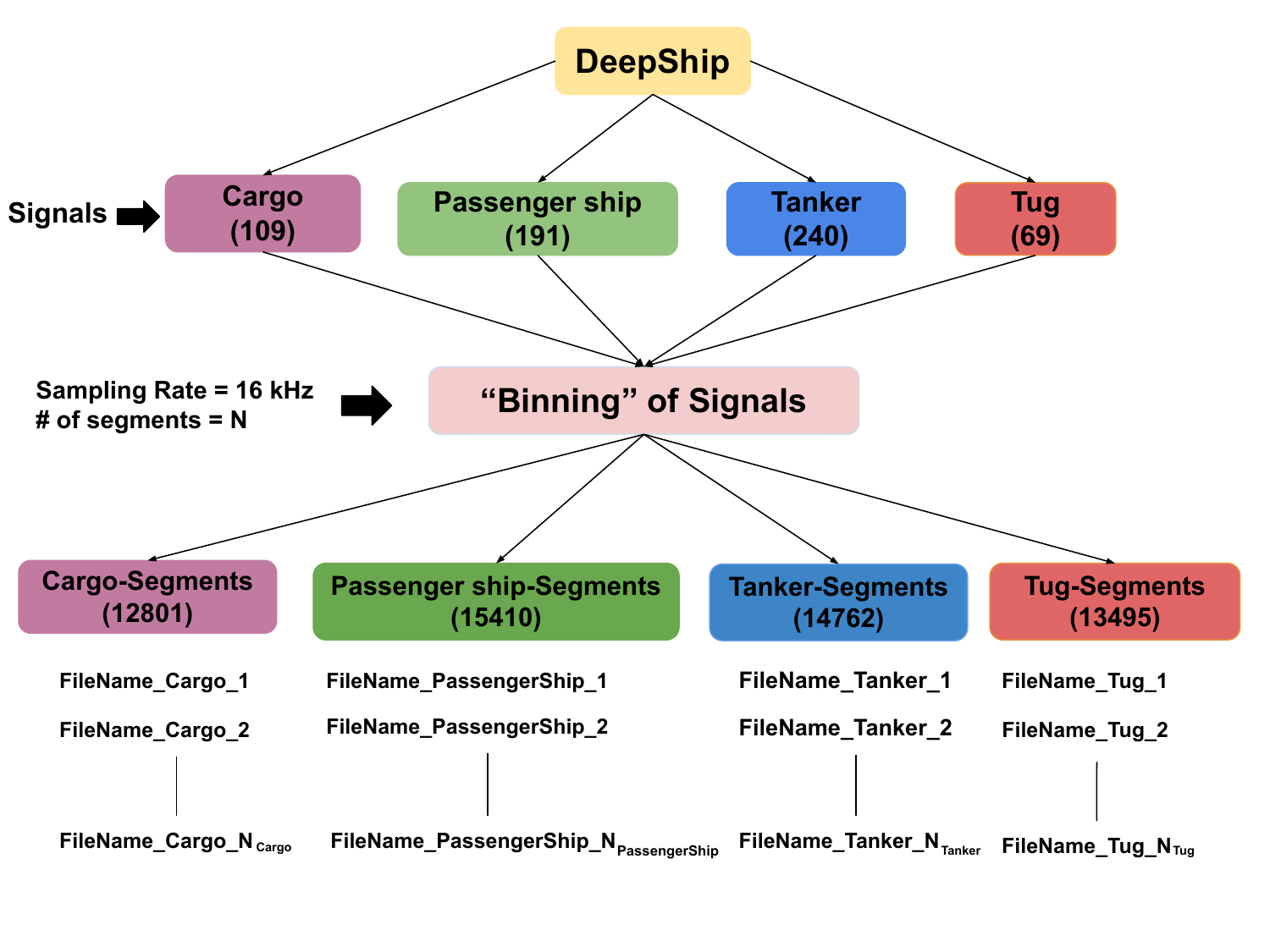}
    \caption{DeepShip dataset structure.}
    \label{fig:systemmodel}
\end{figure}

The DeepShip dataset \cite{irfan2021deepship} was used in this work. The database contained 609 records reflecting the sounds of four different ship types: cargo, passengership, tanker, and tug. Following \cite{irfan2021deepship}, each signal is re-sampled to a frequency of $16$ kHz and divided into segments of three seconds. Figure \ref{fig:systemmodel} illustrates the structure of the dataset after ``binning" the signals into segments. The number of signals and segments for each class are also shown.

\subsection{Experimental Design}

\begin{table*}[htb]
\caption{Overall performance metrics for baseline TDNN and proposed HLTDNN model. The average score with $\pm 1\sigma$ across the three experimental runs of random initialization is shown and the best average metric is bolded. The log of the Fisher Discriminant Ratio (FDR) is shown due to the magnitude of the FDR score. The time-frequency features in this work were Mel Spectrogram (MS), Mel-frequency cepstral coefficients (MFCC), Short-time Fourier transform (STFT), Gammatone-frequency cepstral coefficients (GFCC), Constant-q transform (CQT), and Variable-q transform (VQT).}
\centering
\begin{tabular}{|c|c|c|c|c|c|c|c|}
\hline
Features                        & Model  & Accuracy          & Precision         & Recall          & F1 Score         & MCC              & FDR             \\ \hline
\multirow{2}{*}{\makecell{MS}} & TDNN   &50.31 $\pm$ 1.41\% &39.56 $\pm$ 0.05\% & 47.67 $\pm$  0.03\% &42.09 $\pm$  0.02\%  &34.22 $\pm$  0.02\% & 4.14  $\pm$  1.50 \\ \cline{2-8} 
                                & HLTDNN &47.46 $\pm$ 2.39\% &45.25 $\pm$ 0.03\% &51.80 $\pm$ 0.04\% &46.00 $\pm$  0.03\% &29.55 $\pm$  0.03\% &\textbf{20.51  $\pm$ 1.86}\\ \hline
\multirow{2}{*}{MFCC}           & TDNN   & 51.39 $\pm$ 0.79\% & 50.10 $\pm$ 0.02\%  & 49.95 $\pm$ 0.03\%  & 49.48 $\pm$ 0.02\%  & 34.84 $\pm$  0.01\%  &5.34 $\pm$ 1.29 \\ \cline{2-8} 
                                & HLTDNN &54.41 $\pm$ 0.42\% &54.28 $\pm$ 0.03\%  &53.91 $\pm$ 0.03\% &\textbf{53.62 $\pm$  0.02\%} &39.38 $\pm$ 0.02\% &15.29 $\pm$ 1.85 \\ \hline
\multirow{2}{*}{STFT}           & TDNN   &51.15 $\pm$ 0.72\% &40.88 $\pm$ 0.03\%  &48.49 $\pm$ 0.01\% & 43.86$\pm$ 0.02\%  &24.04 $\pm$ 0.04\% &8.30 $\pm$ 2.87\\ \cline{2-8} 
                                & HLTDNN & \textbf{59.21 $\pm$ 0.56\%} & \textbf{54.84 $\pm$ 0.02\%} &\textbf{56.59 $\pm$ 0.03\% }  & 53.23 $\pm$ 0.02\% & \textbf{46.05 $\pm$  0.01\%} & 17.75 $\pm$  0.58 \\ \hline
\multirow{2}{*}{GFCC}           & TDNN   & 27.73 $\pm$  0.18\%  & 17.45 $\pm$  0.00\% & 26.40 $\pm$  0.00\% & 17.61 $\pm$  0.00\%  &3.63  $\pm$  0.00\% & 15.26  $\pm$ 0.44 \\ \cline{2-8} 
                                & HLTDNN & 43.42 $\pm$  0.61\% & 39.63 $\pm$  0.01\% & 41.44 $\pm$  0.01\% & 38.57 $\pm$  0.01\% & 24.24  $\pm$ 0.01\% & 11.94  $\pm$ 4.82 \\ \hline
\multirow{2}{*}{CQT}            & TDNN   & 36.89 $\pm$   0.83\%  & 23.34 $\pm$   0.03\% & 34.92 $\pm$   0.07\% & 30.85 $\pm$   0.02\% & 15.06 $\pm$   0.01\%&16.95 $\pm$  0.56 \\ \cline{2-8} 
                                & HLTDNN & 50.66 $\pm$ 1.37\% & 44.37 $\pm$   0.01\%  &48.04 $\pm$   0.02\% & 43.62 $\pm$   0.02\% & 34.30 $\pm$   0.02\% & 13.14 $\pm$   3.61 \\ \hline
\multirow{2}{*}{VQT}            & TDNN   &36.76 $\pm$   0.96\%  &28.14 $\pm$   0.02\%  &34.80 $\pm$   0.07\% &30.76 $\pm$   0.02\% &14.84 $\pm$   0.01\% &16.82 $\pm$   0.94 \\ \cline{2-8} 
                                & HLTDNN &50.12 $\pm$   0.27\%  &43.35 $\pm$ 0.02\%  &47.57 $\pm$   0.01\% &43.40 $\pm$   0.01\% &33.44 $\pm$   0.00\% &13.28 $\pm$ 2.87 \\ \hline
\end{tabular}
\label{Tab:mat}
\end{table*}

\noindent\textbf{Feature Extraction} Six different features are extracted: Mel Spectrogram (MS), Mel-frequency cepstral coefficients (MFCCs), Short-time Fourier transform (STFT), Gammatone-frequency cepstral coefficients (GFCC), Constant-q transform (CQT), and Variable-q transform (VQT). The window and hop length for each feature was set to $250$ and $64$ ms respectively \cite{irfan2021deepship}. {The number of Mel filter banks for the MelSpectrogram was set to $40$. For MFCC, the number of Mel-frequency ceptral coeffiencts was $16$. The number of frequency bins for STFT was $48$ while GFCC, CQT, and VQT used 64 frequency bins.} 
The feature dimensions after zero-padding were $48 \times 48$ for MS and STFT, $16 \times 48$ for MFCC, and $64 \times 48$ for GFCC, CQT, and VQT.

\noindent\textbf{Data partitioning} The data set was split into 70\% training, 15\% validation, and 15\% test based on the signals (428 training, 90 validation, and 91 test). After ``binning" the signals into three second segments, 56,468 segments were created (38,523 training, 9,065 validation, and 8,880 test). All segments of each signal remained in the same partition to prevent data leakage (\textit{i.e.}, if one signal was selected for training, all segments of the signal were also used for training).

\noindent\textbf{Experimental setup} The models (TDNN or HLTDNN) were evaluated with each individual feature across three runs of random initialization. The experimental parameters for the models were the following: 
\begin{itemize}
\item {Optimizer:} Adagrad
\item {Learning rate ($\eta$):} 0.001
\item {Batch size:} 128
\item {Epochs:} 100
\item {Dropout ($p$):} 0.5
\item {Early stopping:} 10 epochs
\item {Number of bins (HLTDNN):} 16
\end{itemize}
\noindent Dropout was added before the output classification layer and early stopping was used to terminate training if validation loss did not improve within number of patience epochs. Experiments were conducted on an NVIDIA RTX 3090. The models are implemented in Pytorch 1.13, TorchAudio 2.0, and nnAudio 0.3.1 \cite{cheuk2020nnAudio}.

\section{Results and Discussion}
\subsection{Classification Performance}

\begin{figure}[htb]
  \centering
  \begin{subfigure}{0.235\textwidth}
  \centering
    \includegraphics[width=\textwidth]{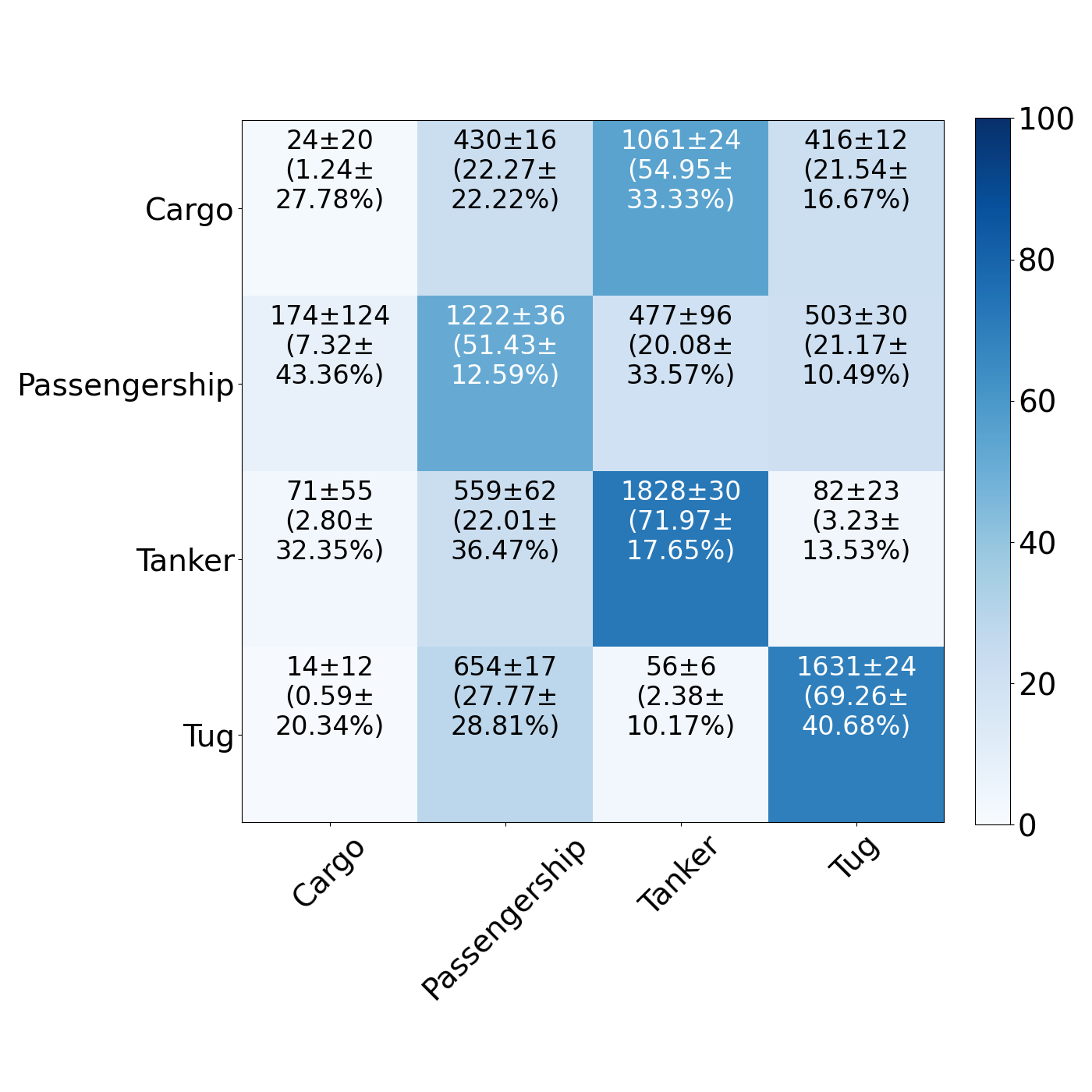}
    \caption{TDNN ($51.15 \pm 0.72\%$)}
    \label{fig:TDNN_CM}
  \end{subfigure}
  \begin{subfigure}{0.235\textwidth}
  \centering
    \includegraphics[width=\textwidth]{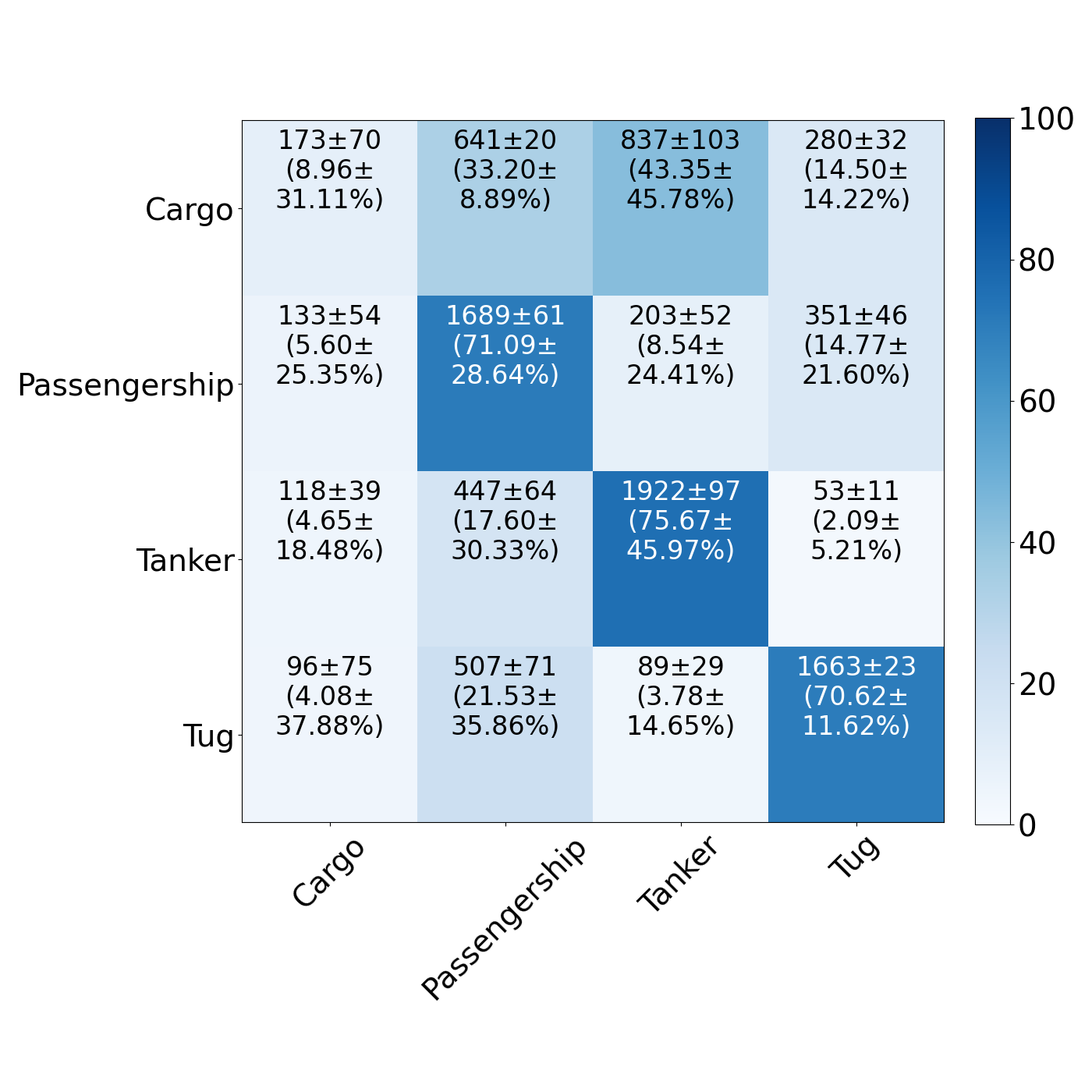}
    \caption{HLTDNN ($59.21 \pm 0.56\%$)}
    \label{fig:HLTDNN_CM}
  \end{subfigure}
\caption{Average confusion matrices for the TDNN and HLTDNN on the DeepShip dataset using the STFT feature. The average overall test accuracy is shown in parenthesis.}
  \label{fig:CM}
  \end{figure}
  
TDNN and HLTDNN classification performances are shown in Table \ref{Tab:mat}. Classification performance was accessed using five metrics: accuracy, precision, recall, F1 score, and Matthew's correlation coefficient (MCC). Fisher's discriminant ratio (FDR) was used to access the feature quality (discussed more in Section \ref{sec:feats}). Confusion matrices for the TDNN and HLTDNN using best performing feature are displayed in Figures \ref{fig:TDNN_CM} and \ref{fig:HLTDNN_CM} respectively. For the HLTDNN, STFT achieved the best classification performance compared to other features. However, MFCC had the best for performance for TDNN across the different performance metrics. STFT performed similarly to MFCC when observing classification accuracy. Additional quantitative and qualitative analysis will use STFT to evaluate the impact of the histogram layer on the vessel classification.

The TDNN model initially performed well with the Mel spectrogram, MFCC, and STFT, but significantly degraded for the other three features (Table \ref{Tab:mat}). The best performance was achieved using the MFCC feature as input while the worst feature was GFCC. A possible reason for this is that each feature used a 250 ms window and hop length of 64 ms. {The short time frame may be limiting the frequency domain and selecting the best frequency band greatly impacts performance {\cite{pollara2016improvement}}.} However, the performance of the HLTDNN was fairly robust across the different time-frequency features. The STFT feature performed the best for this model, and the HLTDNN also improved the performance of the GFCC, CQT and VQT features significantly in comparison to the TDNN. This demonstrates that the statistical context captured by the histogram layer is useful for improving target classification.

Both models did not identify the Cargo class as well as the other vessel types as shown in Figure \ref{fig:CM}. Particularly, the most common classification mistakes occurred when the model predicted Cargo as Tanker (\textit{i.e.}, false positive). Intuitively, this classification error makes sense because Tanker is a type of Cargo ship (\textit{e.g.}, oil tanker \cite{wang2013novel}) and the sound produced by each ship maybe similar. Also, the Cargo class in the DeepShip data has been noted to have high intra-class variance \cite{nie2023contrastive}. As a result, the Cargo class was the most difficult to classify. Feature regularization methods (\textit{i.e.}, constrastive learning) can be incorporated into the objective function to mitigate intra-class variance.  
    
\subsection{Feature Evaluation} \label{sec:feats}

\begin{table}[htb]
\centering
\caption{STFT Fisher's discriminant ratio (FDR) scores for each class and overall. The average score with $\pm 1\sigma$ across the three experimental runs of random initialization is shown and the best average metric is bolded. The log of the FDR is shown due to the magnitude of the FDR score. The higher FDR score indicates better separability and compactness of the features in higher dimensional space for each class.}
\begin{tabular}{ccc}   
& \multicolumn{1}{l}{}                   & \multicolumn{1}{l}{}                            \\ \hline
\multicolumn{1}{|c|}{Class}         & \multicolumn{1}{c|}{TDNN}              & \multicolumn{1}{c|}{HLTDNN}                     \\ \hline
\multicolumn{1}{|c|}{Cargo}         & \multicolumn{1}{c|}{6.00$\pm$4.11}    & \multicolumn{1}{c|}{\textbf{23.65$\pm$7.42}}     \\ \hline
\multicolumn{1}{|c|}{Passengership} & \multicolumn{1}{c|}{6.36$\pm$3.01}     & \multicolumn{1}{c|}{\textbf{19.44$\pm$ 2.56}}   \\ \hline
\multicolumn{1}{|c|}{Tanker}        & \multicolumn{1}{c|}{5.08$\pm$5.75}     & \multicolumn{1}{c|}{\textbf{19.67$\pm$3.11}}  \\ \hline
\multicolumn{1}{|c|}{Tug}           & \multicolumn{1}{c|}{13.01$\pm$1.89} & \multicolumn{1}{c|}{\textbf{20.69$\pm$  3.08}} \\ \hline
\multicolumn{1}{|c|}{Overall}         & \multicolumn{1}{c|}{ 8.30$\pm$2.87}  & \multicolumn{1}{c|}{\textbf{  17.75$\pm$ 0.58}}  \\ \hline
\end{tabular}
\label{tab:FDR}
\end{table}

\begin{figure}[htb]
  \centering
  \begin{subfigure}{0.2365\textwidth}
    \includegraphics[width=\textwidth]{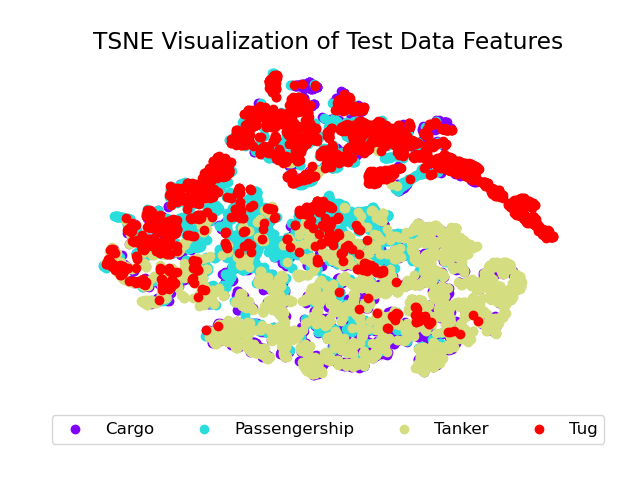}
    \caption{TDNN ($8.30 \pm 2.87$)}
    \label{fig:tesn-tdnn}
  \end{subfigure}
  \begin{subfigure}{0.2365\textwidth}
    \includegraphics[width=\textwidth]{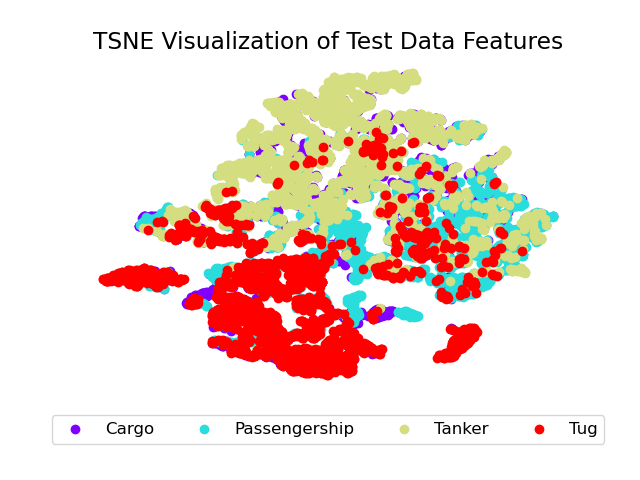}
    \caption{HLTDNN ($17.75 \pm 0.58$)}
    \label{fig:tsne-hltdnn}
  \end{subfigure}
\caption{2D t-SNE projections of features from penultimate layer of the best performing TDNN and HLTDNN on the DeepShip dataset using the STFT feature. Each t-SNE projection used the same initialization for a fair qualitative comparison between the features of each model. The average overall log FDR is shown in parenthesis. The higher FDR score indicates better separability and compactness of the features in higher dimensional space.}
\label{fig:TSNE}
  \end{figure}

In addition to the classification metrics, quality of the features was accessed using Fisher's Discriminant Ratio (FDR). FDR is the ratio of the inter-class separability and the intra-class compactness. Ideally, the inter-class separability should be maximized (\textit{i.e.}, different vessel types should be ``far away" from one another or have large distances between the classes in the feature space) and the intra-class compactness should be minimized (\textit{i.e.}, samples from the same class should be ``close" or have small distances between one another in the feature space). As a result, the FDR should be maximized. From Table \ref{Tab:mat}, the log of the FDR shows that the histogram model achieved the best FDR scores for all six features further demonstrating the utility of the statistical features. 

A deeper analysis using the best performing feature (STFT) in terms of classification performance is shown in Table \ref{tab:FDR}. For all four classes, the log FDR for the HLTDNN is statistically significant (no overlapping error bars) in comparison to the TDNN. The main difference between the two models were the increased feature separability of the HLTDNN model in comparison with the baseline TDNN.  The TDNN had smaller denominator (\textit{i.e.}, intra-class compactness) compared to the HLTDNN when computing the norm of the within-scatter matrix, indicating that the TDNN performs marginally better in terms of intra-class compactness. On the other hand, the features from the HLTDNN are more separable than those from the TDNN, as evident from the norm of the between-scatter matrix, showing the HLTDNN's superiority in terms of inter-class separability. The FDR scores further elucidate the importance of statistical texture information captured by the histogram layer. 

Figure \ref{fig:TSNE} shows the 2D t-SNE projection of the features from the best performing models using the STFT feature. The same random initialization for t-SNE was used for both methods in order to do a fair comparison between both models. The qualitative results of t-SNE  match our quantitative analysis using FDR. 
 The features extracted by the histogram acts as a similarity measure for the statistics of the data and assigning higher ``votes" to bins where features are closer. The addition of these features to the TDNN model improved the separability of the classes as observed in Figure \ref{fig:tsne-hltdnn}. Modifying the histogram layer to help improve the intra-class compactness of the HLTDNN would be of interest in future investigations.

\section{Conclusion}
In this work, a novel HLTDNN model was developed to incorporate statistical information for improved target classification in passive sonar. In comparison to the base TDNN, the HLTDNN not only improved classification performance and led to improved feature representations for the vessel types. Future work will investigate combining features as opposed to using a single time-frequency representation as the input to the network. Each feature can also be tuned (\textit{e.g.}, change number of frequency bins) to enhance the representation of the signals. Additionally, both architectures can be improved by a) adding more depth and b) leveraging pretrained models. The training strategies could also use approaches to mitigate overfitting and improve performance, such as regularization of the histogram layer (\textit{e.g.}, add constraints to the bin centers and widths) and data augmentation. 

\label{sec:conclusion}

\bibliographystyle{IEEEtran}
\bibliography{refs23}

\end{document}